\documentclass[aps,prd,preprintnumbers,superscriptaddress,nofootinbib,showpacs]{revtex4}
\usepackage[dvips]{graphicx}
\usepackage{bm,latexsym,amsmath,amssymb,amsfonts,mathrsfs}
\usepackage{color}
\input{colordvi.tex}
\usepackage[normalem]{ulem}

\newcommand*{\barS}{\widetilde{S}}
\begin{document}

\title{Dynamical breaking of {\it shift-symmetry} in supergravity-based inflation}

\author{Anupam~Mazumdar} 
\email[Email: ]{a.mazumdar"at"lancaster.ac.uk} 
\affiliation{Consortium for Fundamental Physics, Physics Department,
Lancaster University, LA1 4YB, UK}

\author{Toshifumi~Noumi}
\email[Email: ]{toshifumi.noumi``at''riken.jp}
\affiliation{Mathematical Physics Laboratory, RIKEN Nishina Center,
Saitama 351-0198, JAPAN}

\author{Masahide~Yamaguchi}
\email[Email: ]{gucci"at"phys.titech.ac.jp}
\affiliation{Department of Physics, Tokyo Institute of Technology, Tokyo
152-8551, Japan }

\begin{abstract}

{\it Shift-symmetry} is essential to protect the flatness of the
potential, even beyond the super-Planckian vacuum expectation value
(VEV) for an inflaton field. The breaking of the {\it shift-symmetry}
can yield potentials suitable for super-Planckian excursion of the
inflaton.
The aim of this paper is to illustrate that it is indeed
possible to break the {\it shift-symmetry} dynamically within $4$
dimensional supergravity prior to a long phase of inflation.
Thanks to the {\it shift-symmetry},
the leading contribution to the inflaton
potential is free from the dangerous exponential factor
even after its
breaking,
which is the main obstacle to realizing the super-Planckian
inflation in supergravity.
But,
in our simple model,
the resulting inflaton potential is a cosine type potential rather than the power-law one and it is difficult to realize a super-Planckian breaking scale unfortunately.
\end{abstract}

\preprint{RIKEN-MP-89}
\maketitle

\section{Introduction}

The observations of the cosmic microwave background (CMB) temperature
anisotropies~\cite{Hinshaw:2012aka,Ade:2013uln} now strongly support the
occurrence of primordial inflation~\cite{inflation} in the early
Universe. The observed temperature anisotropy can be well fitted by the
primordial perturbations generated during inflation and the
anti-correlation of the temperature (T) and E-mode polarization at large
angular scale suggests that the primordial perturbations have been
stretched on superhorizon scales \cite{Hinshaw:2012aka,Ade:2013uln}. In
addition, very recently, BICEP2 reported the detection of the primordial
tensor perturbations through the B-mode polarization as
\cite{Ade:2014xna}
\begin{equation}
  r =0.20^{+0.07}_{-0.05}\,\, (68\% {\rm CL}),
\end{equation}
where $r$ is the tensor-scalar ratio. To explain this large
tensor-to-scalar ratio is challenging for cosmology and particle physics
because of the Lyth bound \cite{Lyth:1996im}, one would expect a
super-Planckian excursion of the inflaton field in order to generate
large $r$. Of course, the current data can also be explained by the
sub-Planckian excursion of the inflaton field
\cite{Choudhury:2013iaa,Choudhury:2014kma}, or via {\it assisted
inflation}~\cite{Liddle:1998jc} with many copies of the inflaton
field~\cite{Kanti:1999ie}, where the field displacement $\Delta \phi
\simeq0.1M_p\leq M_p$, where $\phi$ is the inflaton and $M_{p} \simeq
2.44 \times 10^{18}$ GeV, but here in this paper we are interested in
studying the opposite limit, when $\Delta \phi > M_{p}$.

Generally speaking, the super-Planckian excursion of the inflaton is
problematic from the effective field theory (EFT) point of view of
particle physics and string theory~\cite{Linde:2005dd}. In particular,
within string theory there are many scales, the string scale, $M_s$, the
compactification scale, $M_c$ and the derived $4$ dimensional Planck
scale, with a spectrum, $ M_s\leq M_c\leq M_p$. Beyond $M_s$ there are
quantum corrections not only to the inflaton potential but also to
the inflaton kinetic term which can lead to various complications,
see~\cite{Chialva:2014rla}. One would require a full non-perturbative
completion of gravity, which we lack sorely within string theory as
well. Even if we assume that we have only one fundamental scale, such as
$M_p$, there are many issues pertaining to the validity of an EFT when
the field's VEV goes beyond $M_p$. In principle, a gauge singlet
inflaton can couple to many degrees of freedom, including the Standard
Model and the hidden sector degrees of freedom,
see~\cite{Mazumdar:2010sa}.  Typically, the individual inflaton's
couplings to matter has to be smaller than $10^{-3}$ to maintain the
flatness of the inflaton potential and also to match the density
perturbations created during inflation. Of course there is no
fundamental justification to make such couplings smaller other than
matching the current constraints arising from the CMB.  

Furthermore, there are higher derivative corrections to the inflaton
kinetic term, see~\cite{Chialva:2014rla}, if we do not take all {\it
infinite} higher derivative terms into consideration, there are
potential problems with {\it ghosts} and quantum instability during
inflation. One cannot ignore the higher derivative terms, because a
priori one does not know what should be the inflaton's kinetic energy,
i.e. the inflaton need not slow rolling throughout the phase of
inflation~\cite{Chialva:2014rla}.

In spite of all these challenges, we wish to ask the question - whether
can we explain {\it at least} a such small inflaton couplings to matter,
and large inflaton's VEV during inflation within an EFT approach by
invoking some symmetry such as {\it shift-symmetry}. Within EFT, one has
to ensure that the inflaton's and {\it all other field's} kinetic terms
are small, and here we simply assume so in some patch of the
Universe just to be within the EFT regime~\cite{Burgess:2014lwa}, though
this still relies on an {\it anthropic arguments}.

In principle, one could imagine a {\it shift-symmetry} as a fundamental
symmetry of nature, which would forbid masses and couplings to an
inflaton field. Such a {\it shift-symmetry} has been for the first time
introduced in the context of chaotic inflation in supergravity (SUGRA)
\cite{Kawasaki:2000yn,Kawasaki:2000ws}. However, based on the same token
if {\it shift-symmetry} remains unbroken inflation would never occur in
our patch of the universe. The {\it shift-symmetry} has to be broken,
but in such a way that the breaking remains {\it soft}, which could be
understood via some dynamics of the fields.  A hard breaking can be
introduced~\cite{Kawasaki:2000yn,Kawasaki:2000ws}, but the predictions
can be lost or one has to resort to some anthropic arguments.

The purpose of this paper is to illustrate a concrete model of dynamical
shift symmetry breaking.  Our model is described within $4$ dimensional
${\cal N}=1$ SUGRA setup and the effective inflaton potential results in
cosine type potential~\cite{Freese:1990rb} without the dangerous
exponential factor. Unfortunately, in our simple model, it is difficult
to realize the super-Planckian breaking scale like natural
inflation.~\footnote{See Refs. \cite{natural,Li:2014xna} for recent
works on natural inflation. Also see
e.g. Refs. \cite{Mazumdar:2010sa,Yamaguchi:2011kg} for other inflation
models in supergravity.}

The organization of the paper is as follows. In the next section, we
will introduce {\it shift-symmetry}, then we will construct a simple
scenario of dynamically breaking of the {\it shift-symmetry} in SUGRA
and explain how it works. In the final section, we will give our
conclusions and discussions.

\section{A brief discussion on {\it shift-symmetry}}

Let us explain how the {\it shift-symmetry} allows the super-Planckian
variation of the inflaton field. Note that this argument is not confined
to a supersymmetric (SUSY) theory but applies to a non-SUSY theory.  A
{\it shift-symmetry} is characterized by a symmetry under the following
transformation of a (real) inflaton field $\phi$,
\begin{equation}
  \phi \rightarrow \phi + c \quad (c : {\rm real~~constant}).
\end{equation}
As long as this symmetry is exact, the potential of the inflaton is
completely flat and any field variation even beyond the reduced Planck
scale $M_{p}$ is allowed. This is an essential idea. However, inflation
must end to reheat our Universe, then the {\it shift-symmetry} must be broken
to generate the gradient of the potential.

As far as we know, in all of the models considered so far, the {\it
shift-symmetry} is broken simply by hand or by introducing an auxiliary
field, spurion field, with no kinetic term, whose non-zero VEV is given
by hand.  For example, in SUGRA models, it is often assumed that the
K\"ahler potential respects the {\it shift-symmetry} while the
superpotential breaks the {\it shift-symmetry}.  In such a case, any
kind of superpotential can appear because there is no founding principle
behind the breaking of {\it shift-symmetry}. The introduction of a
spurion field might cure such ambiguity because the original action
before giving a non-zero VEV to the spurion field respects the {\it
shift-symmetry} in this approach. Then, the interactions, or the forms
of the K\"ahler potential and the superpotential, can be
constrained. For example, let us introduce a spurion field $S$ and
extend the {\it shift-symmetry} as \cite{Kawasaki:2000ws}
\begin{eqnarray}
  \phi &\rightarrow& \phi + c \quad (c : {\rm real~~constant}), \\
  S &\rightarrow& S \frac{\phi}{\phi+c}.
\end{eqnarray}
Then, the combination $S\phi$ is invariant under the {\it
shift-symmetry}.  Once this spurion field $S$ takes a non-zero VEV,
i.e. $\langle S \rangle = m$, the {\it shift-symmetry} is broken and the
potential is generated. The key points are, that

\begin{itemize}

\item{the inflaton field $\phi$ always appears in the combination:
$\langle S \rangle \phi = m\phi$, where $m\ll M_p$.  As long as
$m\phi\leq M_p^2$, the EFT treatment is still justified, in spite of the
fact that the cutoff scale of the inflaton is now raised to $M_p^2 /
m$.}

\item{no super-Planckian masses of fermions and bosons appear, because
any interactions of the inflaton including Yukawa and four-point
interactions are suppressed by the small scale $m\ll M_p$.}

\item{if we take the $m \rightarrow 0$ limit, the {\it shift-symmetry}
is restored. In this sense, this model is technically natural. Thus,
chaotic inflation can be naturally realized in this setup and the model
given in Refs. \cite{Kawasaki:2000yn,Kawasaki:2000ws} is a concrete
realization in the context of SUGRA.}

\end{itemize}

However, even in this setup, the non-zero VEV of the field $S$, i.e.
the breaking of the {\it shift-symmetry} has been introduced by hand
unfortunately, by assuming that it is a spurion field. Needless to say,
it is better to break the {\it shift-symmetry} dynamically because,
otherwise, we cannot control the whole dynamics of the system, or
evaluate the effects of the {\it shift-symmetry} breaking adequately. In
this paper, we address this issue and propose a concrete model of the
dynamical breaking of the shift symmetry in SUGRA.

\section{Dynamical breaking of {\it shift-symmetry}.}

In this section, we are going to construct a concrete model of dynamical
breaking of the {\it shift-symmetry} in ${\cal N}=1$ SUGRA.  The key
observation is that the following superpotential,
\begin{equation}\label{supot-1}
  W = e^{a \Phi}
\end{equation}
is invariant (up to a constant phase) under the {\it shift-symmetry},
\begin{equation}
  \Phi \rightarrow \Phi + i \frac{C}{a},
\end{equation}
where $a$ and $C$ are real constants. In fact, the scalar potential in
the global SUSY limit is given by
\begin{equation}
  V(\Phi) = a^2 e^{a(\Phi + \Phi^{\ast})},
\end{equation}
which depends only on the real part of $\Phi$. Thus, the {\it
shift-symmetry} on the imaginary part of $\Phi$ remains\,\footnote{From
here onwards, we denote the scalar components of the superfields by the
same symbols as the corresponding superfields.}. On the other hand, the
following superpotential,
\begin{equation}\label{supot-2}
  W = e^{a \Phi} + e^{-a \Phi}
\end{equation}
is {\it not} invariant under the {\it shift-symmetry}. In fact, the scalar
potential in the global SUSY limit is given by
\begin{equation}
  V(\phi,\chi) = a^2 \left[ e^{\sqrt{2}a \chi}  + e^{-\sqrt{2}a \chi}
           - 2 \cos{\left( \sqrt{2}a \phi \right)}
  \right],
\end{equation}
where 
\begin{equation}
  \Phi = \frac{1}{\sqrt{2}} \left( \chi + i \phi \right).
\end{equation}
It should be noticed that the scalar potential depends not only on the
real part of $\Phi$, i.e. $\chi$, but also on the imaginary part of
$\Phi$, i.e.  $\phi$. In order to recover the {\it shift-symmetry} for
the second type of the superpotential, see Eq.~(\ref{supot-2}), we need
to introduce a pair of superfields, $S$ and $\barS$, and another
superfield $X$\,\footnote{The superfield $X$ is not necessary only for
recovery of the {\it shift-symmetry}. It is also useful to
guarantee the positivity of the potential during inflation
\cite{Kawasaki:2000yn,Kawasaki:2000ws,Kallosh:2010xz}.}. Let us now
consider the following superpotential,
\begin{equation}\label{inflaton-supot}
 W_{I} = v \left( S^n e^{a \Phi} + \barS^n e^{-a \Phi} \right) X,
\end{equation}
where $v \ll 1$ is a constant (in Planck units), and $n$ is a positive
integer number. This superpotential is invariant under the following
{\it shift-symmetry},
\begin{eqnarray}
  \Phi &\rightarrow& \Phi + i \frac{n C}{a}, \nonumber \\
  S &\rightarrow& S e^{-iC}, \nonumber \\
  \barS &\rightarrow& \barS e^{iC}, \nonumber \\
  X &\rightarrow& X. 
\label{eq:shift} 
\end{eqnarray}
However, in order to realize inflation, this {\it shift-symmetry} must
be broken. For this purpose, we introduce another superfield $T$ and add
the following superpotential,
\begin{equation}
  W_{B} = \lambda T \left( S \barS - \mu^2 \right),
\end{equation}
where $\lambda\leq {\cal O}(1)$ and $\mu\leq {\cal O}(1)$ are
constants. Then, the total superpotential, which is given by:
\begin{equation}
 W = W_I + W_B\,,
\end{equation}
is invariant under the {\it shift-symmetry}, Eq.~(\ref{eq:shift}), along with 
\begin{equation}
 T \rightarrow T. 
\label{eq:shift2} 
\end{equation}
One can easily understand that, once the scalar components of the
superfields $\langle S\rangle \neq 0$ and $\langle \barS\rangle
\neq 0$, acquire non-zero VEVs, the {\it shift-symmetry} is broken
dynamically. Let us consider the following K\"ahler potential of the
type\,\footnote{The linear term of $\Phi + \Phi^{\ast}$ can appear
in the K\"ahler potential because of the absence of the $Z_2$
symmetry. Such an effect causes two effects. First one is additional
contribution to the D-term. Second one is a slight deviation of the
minimum of $\chi$ field during inflation from the global minimum. This
deviation is still compatible with the D-flat condition because its
deviation exactly cancels out the additional contribution to the
D-term. So, the essential dynamics remains unchanged and we omit it for
simplicity.}:
\begin{equation}
 K = \frac12 \left(\Phi + \Phi^{\ast} \right)^2 
     + |S|^2 +|\barS|^2 +|T|^2 + |X|^2,
     \label{Kahler}
\end{equation}
which is invariant under the {\it shift-symmetry}, Eqs.~(\ref{eq:shift})
and (\ref{eq:shift2}), and generates the canonical kinetic terms for all
of the fields. Note that {\it shift-symmetry} will also allow
higher order terms, such as
$(\Phi+\Phi^{\ast})^4,~(\Phi+\Phi^{\ast})^6,\cdots, |S|^4,~|S|^6,\cdots,
|\bar S|^4,~|\bar S|^6,\cdots, |T|^4,~|T|^6,\cdots, |X|^4,~|X|^6\cdots,
S^n e^{a\Phi}, \barS^n e^{-a\Phi},\cdots$,
etc., where $\cdots$ contain higher order terms to all {\it infinite}
orders. These terms will give corrections to the canonical kinetic
terms. But, as long as $(\Phi+\Phi^{\ast}), |S|, |\barS|, \cdots \ll
1$, which can be realized dynamically in our model, these corrections
are negligible. Of course, at initial period, we assume the presence of
at least one patch of the Universe, in which the kinetic energies of
{\it all} of the fields are smaller than the Planck energy density and
subdominant.

The Higher derivative terms like ${\cal D}_a \Phi {\cal D}^a \Phi^{\ast}$ in the K\"ahler potential are
also allowed from our symmetry, see \cite{Baumann:2011nm}.
Unless these higher order terms are suppressed by $(\Phi+\Phi^{\ast})^2$ for example,
the derivative expansion may not be justified because of the super-Planckian value of $\phi$
(see again Ref.~\cite{Chialva:2014rla}). One would need to take all {\it infinite} higher derivative corrections
in order to avoid {\it ghosts} and instability of the vacuum~\cite{Chialva:2014rla}. This would require a complete
ultraviolet completion of inflaton and gravitational sector, which we do not aim to address in this paper. Instead,
we make an assumption that the inflationary patch is always within an EFT regime.

Further note that the present model possesses $U(1)_{\rm R}$
symmetry, under which
\begin{eqnarray}
    \Phi(\theta) &\rightarrow& \Phi(\theta e^{i\alpha}),
    \nonumber \\ 
    S(\theta) &\rightarrow& S(\theta e^{i\alpha}),
    \nonumber \\ 
    \barS(\theta) &\rightarrow& \barS(\theta e^{i\alpha}),
    \nonumber \\ 
    X(\theta) &\rightarrow& e^{-2i\alpha} X(\theta e^{i\alpha}),
    \nonumber \\ 
    T(\theta) &\rightarrow& e^{-2i\alpha} T(\theta e^{i\alpha}).
  \label{eq:Rsymmetry}
\end{eqnarray}
The scalar potential in ${\cal N}=1$ SUGRA is given by
\begin{eqnarray}
 V &=& e^K \left[ 
    \biggl| va \left( S^n e^{a \Phi} - \barS^n e^{-a \Phi} \right) X
           + \left(\Phi + \Phi^{\ast}\right) W \biggr|^2
  + \biggl| nv S^{n-1} e^{a \Phi}X + \lambda T \barS
           + S^{\ast} W \biggr|^2 \right. \nonumber \\     
   && \qquad+ \biggl| nv \barS^{n-1} e^{-a \Phi}X + \lambda T S
           + \barS^{\ast} W \biggr|^2      
  + \biggl| \lambda \left( S \barS - \mu^2 \right)
           + T^{\ast} W \biggr|^2 \nonumber \\     
   && \qquad \left. 
  + \biggl| v \left( S^n e^{a \Phi} + \barS^n e^{-a \Phi} \right)
           + X^{\ast} W \biggr|^2 
  - 3|W|^2
          \right] 
  \,\,+\,V_D.
\label{eq:scalarp}
\end{eqnarray}
From here onwards we set $M_p=1$. In the above potential, $V_D$,
represents the D-term contribution, which is given by
\begin{equation}\label{Dterm}
 V_D = \frac{e^2}{2} \left( \left| \barS \right|^2 - \left| S \right|^2
       +\frac{n}{a}(\Phi+\Phi^{\ast}) \right)^2,
\end{equation}
with $e$ being a gauge coupling constant. Such a term can be present if
the {\it shift-symmetry} is gauged by changing the constant parameter $C$ to a
spacetime dependent one $C(x)$.

\section{Inflationary potential}

Now, let us take a closer look at the dynamics of this system. First,
let us assume that the energy scale of the {\it shift-symmetry} breaking
sector, i.e. $W_B$, is much higher than that of the inflation sector,
i.e. $W_I$, which requires:
\begin{equation}
  \lambda^2 \mu^4 \gg v^2 \mu^{2n} \quad
   \Longleftrightarrow \quad \lambda \gg v \mu^{n-2}.
\end{equation}
Under this assumption, the potential energy is roughly given by $V
\simeq \lambda^2 \mu^4$ at the onset of inflation, and the Hubble
expansion rate: $ H^2\simeq V/3\simeq \lambda^2\mu^4/3$.  At such higher
energies, hybrid-type inflation~\cite{Dvali:1994ms,Linde:1997sj} can
occur, where $T$ cannot take a value larger than unity (in Planck
unites) due to the exponential factor $e^K$ in the potential, see
Eq.~(\ref{eq:scalarp}). Then, the mass squared of the field $X$ is
estimated to be:
\begin{equation}
  m_{X}^2 \simeq \lambda^2 \mu^4 \left(1+|T|^2\right)
          \simeq 3H^2 \left(1+|T|^2\right),
\end{equation}
which dynamically drives the field $X$ to the zero VEV. It can be easily
confirmed that, even after this inflation, $m_{X}^2$ is always positive,
so that $X$ stays at the origin for ever.  By inserting $X=0$ to the
scalar potential, Eq.~(\ref{eq:scalarp}) yields
\begin{eqnarray}
  V |_{X=0} &=& e^K \left[ 
    \lambda^2 |T|^2 \left(\Phi + \Phi^{\ast}\right)^2 
              \left| S \barS - \mu^2 \right|^2
  + \lambda^2 |T|^2 \left(
          \left| \barS\left(1+|S|^2\right)-\mu^2 S^{\ast}
	  \right|^2 
        + \left| S\left(1+|\barS|^2\right)-\mu^2 \barS^{\ast}
	  \right|^2 \right) 
  \right. \nonumber \\ 
  &&\left. \qquad
  + \lambda^2 \left( 1 - |T|^2 + |T|^4 \right) 
              \left| S \barS - \mu^2 \right|^2
  +v^2 \biggl| S^n e^{a \Phi} + \barS^n e^{-a \Phi} \biggr|^2 
  \right] 
  \,\,+\frac{e^2}{2} \left( \left| \barS \right|^2 - \left| S \right|^2
                            +\frac{n}{a}(\Phi+\Phi^{\ast}) \right)^2\,,
\label{eq:scalarp2}
\end{eqnarray}
and the mass terms for $S$ and $\barS$ are estimated as
\begin{eqnarray}
m^2_{S,~\bar S}   &\simeq& - \lambda^2 \mu^2 
            \left( S\barS + S^{\ast}\barS^{\ast} \right)
          + \lambda^2 |T|^2 \left(1+\mu^4\right)
             \left(|S|^2 + \left|\barS\right|^2 \right)
     \nonumber \\
   &=& \lambda^2 \biggl[ \left(1+\mu^4\right)|T|^2+\mu^2 \biggr] 
         \left|\Psi\right|^2
     + \lambda^2 \biggl[ \left(1+\mu^4\right)|T|^2-\mu^2 \biggr] 
         \left|\overline{\Psi}\right|^2,
  \label{eq:potX}
\end{eqnarray}
where we have defined
\begin{eqnarray}
  \Psi = \frac{1}{\sqrt{2}} \left(S-\barS^{\ast}\right),
    \quad
  \overline{\Psi} = \frac{1}{\sqrt{2}} \left(S+\barS^{\ast}\right),
\end{eqnarray}
and we have taken $n \ge 2$ in Eq.~(\ref{inflaton-supot}). Since
$m_{\Psi}^2 \gg \lambda^2 \mu^4 \simeq 3H^2$, the $\Psi$ field has a
Hubble-induced mass and quickly settles down to the zero VEV within one
Hubble time or so, which implies $S = \barS^{\ast}$ and $|S| =
|\barS|$. This condition is compatible with the D-term flatness
condition, $V_D=0$, along with $\Phi+\Phi^{\ast} = 0$, which holds true
for almost all periods. At this point, we can discuss the dynamics of
the fields for two particular scenarios:

\begin{itemize}

\item {\underline{ $|T| \gtrsim T_c $, {\it dynamically preserving
shift-symmetry} }:\\ As long as the VEV of $T$ is such that : $|T|
\gtrsim T_c \simeq \mu$, or, $m_{\Psi}^2 \gg \lambda^2 \mu^4 \simeq
3H^2$, which also leads dynamically to $\overline{\Psi} = 0$. Therefore,
for $|T| \gtrsim T_c$, $S$ and $\barS$ stay at the origin and the
potential $V$ is dominated by $\lambda^2 \mu^4$, leading to the hybrid
inflation~\cite{Dvali:1994ms,Linde:1997sj}.

The SUGRA effects and the one-loop potential coming from the SUSY
breaking effects could drive the inflaton field $T$ like in the case of
standard hybrid inflation. It should be noticed that, during this
inflation, the effective mass squared of the real part of $\Phi$,
$\chi$, is approximately $3H^2$. Therefore, $\chi
(=(\Phi+\Phi^{\ast})/\sqrt{2})$ quickly rolls down to its minimum, that
is, the zero as well. On the other hand, the imaginary part of $\Phi$,
$\phi$, is still arbitrary. That is, the {\it shift-symmetry} which is
preserved at this stage.}

\item {\underline{$|T| \lesssim T_c$, {\it dynamically breaking
shift-symmetry}}: \\ In this case the effective mass squared
$m_{\overline{\Psi}}^2< 0$, with its magnitude is larger than the Hubble
parameter squared, the $\overline{\Psi}$ field becomes unstable so that
the fields $S$ and $\barS$ quickly roll down to the minimum of
the potential with $S\barS = \mu^2$ and $|S| =
\left|\barS\right|$ together with $\Phi+\Phi^{\ast} = 0$, which
can be parametrized as
\begin{equation}
  S = \mu e^{i\beta}, \quad \barS = \mu e^{-i\beta}
\end{equation}
with $\beta$ being a real constant. Thus, the fields $S$ and
$\barS$ acquire the non-zero VEVs, which dynamically breaks the
{\it shift-symmetry}.}

\end{itemize}

Further note that, for $S\barS \simeq \mu^2$, the
effective mass squared of $T$, $m_{T}^2$, is estimated as
\begin{equation}  
  m_{T}^2 \simeq 2 \lambda^2 \mu^2,
\end{equation}
which mainly comes from the second and third terms in the right hand
side of the first line in Eq. (\ref{eq:scalarp2}). Thus, after the end
of hybrid inflation, $T$ quickly settles down to its minimum,
i.e. $\langle T\rangle =0$. Then, the effective scalar potential with,
$\langle X\rangle =\langle T\rangle =0$, is given by
\begin{eqnarray}
  V |_{X=T=0} &=& e^K \left[ 
        \lambda^2 \left| S \barS - \mu^2 \right|^2
      +v^2 \biggl| S^n e^{a \Phi} + \barS^n e^{-a \Phi} \biggr|^2 
                     \right]
  \,\,+\frac{e^2}{2} \left( \left| \barS \right|^2 - \left| S \right|^2
                     +\frac{n}{a}(\Phi+\Phi^{\ast}) \right)^2
\label{eq:scalarp3}
\end{eqnarray}
with $K = \chi^2 + |S|^2 + \left|\barS\right|^2$. It is manifest
that this effective potential is positive definite and its global
minimum is given by the conditions
\begin{eqnarray}
  && S\barS - \mu^2 = 0, \nonumber \\
  && S^n e^{a \Phi} + \barS^n e^{-a \Phi} = 0, \nonumber \\
  && \left| \barS \right|^2 - \left| S \right|^2
                     +\frac{n}{a}(\Phi+\Phi^{\ast})=0.
\end{eqnarray} 
These conditions lead to the global minimum for the fields, as
\begin{eqnarray}
  S_{\rm min} &=& \mu e^{i\beta}, \\
  \barS_{\rm min} &=& \mu e^{-i\beta}, \\
  \chi_{\rm min} &=& 0, \\ 
  \phi_{\rm min} &=& - \frac{\sqrt{2}n\beta}{a} +
                        \frac{(2m-1)}{\sqrt{2}a}\pi,
\end{eqnarray}
where $m$ being an integer number.

However, when hybrid inflation ends and the {\it shift-symmetry} is
broken with $S\barS=\mu^2$, the imaginary part of $\Phi$ does not
necessarily stay at the minimum, because before the breaking of the {\it
shift-symmetry} all the values of the imaginary part of $\Phi$, $\phi$,
are equally distributed, thanks to the {\it shift-symmetry}. Thus, the
initial condition of $\phi$ is determined accidentally. The effective
potential is given by
\begin{eqnarray}
  V_{\rm eff} &=& e^K v^2 
     \biggl| S_{\rm min}^n e^{a \Phi} + \barS_{\rm min}^n 
          e^{-a \Phi} \biggr|^2, \\
                 &=& e^{\chi^2+2\mu^2} \cdot v^2 \mu^{2n}
     \left[ e^{\sqrt{2}\chi/M} + e^{-\sqrt{2}\chi/M}
           + 2 \cos\left( 2n\beta+\frac{\sqrt{2}\phi}{M} \right) \right],
\label{V_eff}
\end{eqnarray}
with $M = 1/a$. Here, let us identify the inflaton and the
Nambu-Goldstone (NG) boson correctly, which are given by
\begin{eqnarray}
  \phi_{\rm inf} &=& \phi + \frac{nM}{\mu} \beta_c, \\
  \phi_{\rm NG} &=& \phi - \frac{nM}{\mu} \beta_c.
\end{eqnarray}
with $\beta_c \equiv \sqrt{2} \mu \beta$.
Then, the covariant kinetic
terms are given by
\begin{eqnarray}
&&  \frac12 \left( D_{\mu} \phi \right)^2
+ \frac12 \left( D_{\mu} \beta_c \right)^2 \nonumber \\
= &&
\frac12 \left[ 
  \frac14 \left(1+\frac{\mu^2}{n^2 M^2}\right)
  \left\{ \left( \partial_{\mu} \phi_{\rm inf} \right)^2
         +\left( \partial_{\mu} \phi_{\rm NG} \right)^2 \right\}
  +  \left(1-\frac{\mu}{n M}\right)
        \partial_{\mu} \phi_{\rm inf}\partial^{\mu} \phi_{\rm NG} 
 \right.
 \nonumber \\
&& \left.  +\sqrt{2} nM A_{\mu} \left\{
      \left(1-\frac{\mu^2}{n^2 M^2}\right) \partial^{\mu} \phi_{\rm inf}
    + \left(1+\frac{\mu^2}{n^2 M^2}\right) \partial^{\mu} \phi_{\rm NG} 
    \right\}
  + 2 (n^2 M^2 + \mu^2) A_{\mu} A^{\mu}
\right].
\end{eqnarray}
where 
\begin{eqnarray}
 D_{\mu} \phi \equiv \partial_{\mu} \phi + \sqrt{2} n M A_{\mu},\qquad
 D_{\mu} \beta_c \equiv \partial_{\mu} \beta_c - \sqrt{2}\mu A_{\mu},
\end{eqnarray}
with $A_{\mu}$ being the gauge field. The NG boson $\phi_{\rm NG}$ is
eaten by the gauge field, so the remaining kinetic terms in the unitary
gauge become
\begin{eqnarray}
&& \frac12 \left[ 
  \frac14 \left(1+\frac{\mu^2}{n^2 M^2}\right)
  \left( \partial_{\mu} \phi_{\rm inf} \right)^2
  + \sqrt{2}nM \left(1-\frac{\mu^2}{n^2 M^2}\right) A_{\mu} \partial^{\mu} \phi_{\rm inf}
  + 2(n^2 M^2 + \mu^2) A_{\mu} A^{\mu}
\right] \nonumber \\
&& =\frac12 \frac{1}{1+\frac{n^2M^2}{\mu^2}}
    \left( \partial^{\mu} \phi_{\rm inf} \right)^2
  +  (\mu^2 +n^2 M^2) \widetilde{A}_{\mu} \widetilde{A}^{\mu} 
  \nonumber \\
&& = \frac12 \left( \partial^{\mu} \widetilde{\phi}_{\rm inf} \right)^2
  +  (\mu^2 +n^2 M^2) \widetilde{A}_{\mu} \widetilde{A}^{\mu},
\end{eqnarray}
where 
\begin{eqnarray}
&& \widetilde{A}_\mu \equiv A_\mu 
  + \frac{n M-\frac{\mu^2}{n M}}{\sqrt{2}(\mu^2 +n^2 M^2)} 
     \partial_{\mu} \phi_{\rm inf}, \\
&& \widetilde{\phi}_{\rm inf} \equiv 
        \sqrt{\frac{1}{1+\frac{n^2M^2}{\mu^2}}}\phi_{\rm inf}.
\end{eqnarray}
Thus, the effective potential for the canonically normalized inflaton
$\widetilde{\phi}_{\rm inf}$ is given by
\begin{eqnarray}
  V_{\rm eff}(\widetilde{\phi}_{\rm inf}) =
     2 e^{2\mu^2} \cdot v^2 \mu^{2n}
         \cos\left( \frac{\sqrt{2}}{M} \sqrt
      {1+\frac{n^2M^2}{\mu^2}}
    \widetilde{\phi}_{\rm inf} \right),
\label{V_eff}
\end{eqnarray}
where the decay constant $f$ is given by
\begin{eqnarray}
  f = \frac{M}{\sqrt{2}} \,\sqrt{\frac{1}{1+\frac{n^2M^2}{\mu^2}}}
   \rightarrow \frac{\mu}{\sqrt{2}\,n} \quad {\rm for} \quad {nM \gg \mu}.
\end{eqnarray}
Thus, since $n$ is an integer number and larger than unity in this
simple example, the decay constant $f$ cannot be super-Planckian scale
as long as $\mu$ is sub-Planckian scale. So, inflation becomes
hilltop type one.

In order to reheat the Universe after inflation, we introduce the
following superpotential,
\begin{equation}
  W_{R} = y S^n e^{a\Phi} N N,
\end{equation}
where $y$ is a (Yukawa) coupling constant and $N$ is the right-handed
neutrino superfield. This superpotential with the canonical K\"ahler
potential for $N$ is manifestly invariant under the {\it shift-symmetry}
Eqs.~(\ref{eq:shift}), (\ref{eq:shift2}), and $N \rightarrow N$. Once
$S$ acquires the non-zero VEV, this superpotential leads to a Yukawa
coupling between the inflaton $\phi$ and the right-handed neutrino
$\widetilde{N}$. Therefore, the leptogenesis through the inflaton decay
and the reheating of the Universe through the decay of the right handed
neutrino to the standard particles are possible by tuning the parameters
adequately.

\section{Conclusions and discussions}

In this paper, we constructed a concrete example of the dynamical
breaking of the {\it shift-symmetry} in SUGRA. By taking the exponential
type of the superpotential for $\Phi$, which might appear through some
non-perturbative effects, we first consider the superpotential invariant
under the {\it shift-symmetry}. Then, by arranging the GUT Higgs-like
superpotential as well, the {\it shift-symmetry} is dynamically
broken. The inflaton has no dangerous exponential factor at the
leading order in the scalar potential even after the shift symmetry
breaking. Such an exponential factor is the main obstacle to realizing
super-Planckian inflation in supergravity.
Unfortunately, in our simple
model, the potential obtained for the inflaton is a cosine type
potential rather than power-law one and it is difficult to realize a
super-Planckian decay constant. One possible way to obtain the super-Planckian
decay constant with $\mu$ being the sub-Planckian scale is to make $n$
smaller than unity. Of course, $n$ is an integer number and larger than
unity in this simple example. However, for example, if we start from the
higher order K\"ahler potential for $S$ ($\barS$) like $|S|^{2m}$
($|\barS|^{2m}$) instead of the canonical K\"ahler potential
(with $S\barS$ replaced by $(S\barS)^m$ in the
superpotential $W_B$ at the same time), then such a model becomes
equivalent to our simple model with the effective $n_{\rm eff} = n/m$ by
field redefining $S' \equiv S^m (\barS'
\equiv \barS^m)$. Thus, if we take $n_{\rm eff} \lesssim \mu/M_p
\lesssim 1$, the decay constant $f \gtrsim M_{p}$, which may realize
super-Planckian inflation like natural inflation. We leave more
realistic realization of natural and chaotic inflation as a future work.

We have restricted ourselves within the regime of EFT, where the fields
have masses and energy densities below the cut-off in spite of the fact
that the inflaton VEV could be large and above the Planck scale. We have
also pointed out that it is possible to generate small inflaton
couplings to matter in order to avoid some of the quantum corrections to
the inflaton potential~\cite{Chialva:2014rla}. In this paper we have
explicitly assumed that all the fields are slow rolling initially
in some patch of the Universe, such that the kinetic energy is indeed
sub-dominant to be well within the regime of EFT.

\bigskip
{\bf Note added.}  While we finalized the paper, Ref.~\cite{Li:2014xna}
appeared, in which a similar breaking of shift symmetry is given, though
the model is rather different and more complicated. We thank the authors
of~\cite{Li:2014xna} for noticing us that fact.

\acknowledgments
We would like to thank Keisuke Harigaya and Masahiro Ibe for pointing
out the error of the first draft and useful comments and discussions.
A.M. is supported by the Lancaster-Manchester-Sheffield Consortium for
Fundamental Physics under STFC grant ST/J000418/1. The work of T.N. is
supported in part by the Special Postdoctoral Researcher Program at
RIKEN. M.Y. is supported in part by the JSPS Grant-in-Aid for Scientific
Research Nos.~25287054 and 2661006.


\end{document}